\documentclass[aip,pop,preprint,amsmath,amssymb]{revtex4-1}

\usepackage{graphicx}
\usepackage{bm}
\renewcommand{\vec}[1]{\boldsymbol{#1}}

\usepackage{color,ulem} 
\newcommand{\bluetext}[1]{{#1}}

\renewcommand{\div}{\nabla \cdot}
\newcommand{\rot}{\nabla \times}

\def\jgr{{J. Geophys. Res.} }

\def\prl{{Phys. Rev. Lett.} }
\def\pop{{Phys. Plasmas} }

\def\ssr{{Space Science Reviews} }

\begin{document}

\title{Some remarks on the diffusion regions in magnetic reconnection}

\author{Seiji Zenitani}
\email{seiji.zenitani@nao.ac.jp}
\affiliation{Division of Theoretical Astronomy, National Astronomical Observatory of Japan, 2-21-1 Osawa, Mitaka, Tokyo 181-8588, Japan}
\author{Takayuki Umeda}
\affiliation{Solar-Terrestrial Environment Laboratory, Nagoya University, Nagoya, Aichi 464-8601, Japan}

\date{Received 22 December 2013; accepted 14 March 2014; published online 27 March 2014}

\begin{abstract}
The structure of the diffusion regions in antiparallel magnetic reconnection is
investigated by means of a theory and a Vlasov simulation. 
The magnetic diffusion is considered as relaxation to the frozen-in state,
which depends on a reference velocity field.
A field-aligned component of the frozen-in condition is proposed to evaluate
a diffusion-like process. 
Diffusion signatures with respect to ion and electron bulk flows
indicate the ion and electron diffusion regions near the reconnection site. 
The electron diffusion region resembles the energy dissipation region.
These results are favorable to a previous expectation that
an electron-scale dissipation region is surrounded by an ion-scale Hall-physics region.
[http://dx.doi.org/10.1063/1.4869717]
\end{abstract}


\maketitle

Magnetic reconnection relies on
the dissipation mechanism in a narrow region surrounding the X-line,
the so-called diffusion region (DR). 
The structure, dynamics, and observational signatures of the DR
are key topics in reconnection physics,
and therefore
the DR is the main target of
NASA's forthcoming Magnetospheric Multiscale (MMS) mission,\cite{burch09} 
which will probe near-earth reconnection sites at higher resolution than before. 

The word ``diffusion region'' indicates
a region where magnetic diffusion takes place. 
However, the concept of the magnetic diffusion is ambiguous. 
It usually appears in textbooks on magnetohydrodynamics (MHD)
in a reduced form. 
A single MHD fluid is considered in a stationary frame
and then
an electric resistivity is given by the Ohm's law.
Despite these simplifying assumptions,
it is very difficult to formulate the magnetic evolution.\citep{wilmotsmith05}

In collisionless reconnection,
it is expected that the DR develops a nested structure of
an outer ion-physics layer and an inner electron-physics layer
due to different magnetization.\cite{biskamp97,hesse01,drake07}
These layers are popularly called
the ion diffusion region (IDR) and
the electron diffusion region (EDR), 
but they are only loosely defined. 
Signatures of these DRs,
in particular for the EDR,
have been investigated over many years.
\citep{mozer05,shay07,scudder08a,prit09a,zeni11c}
In the case of the EDR,
one of the most promising EDR signatures is
the nonideal energy dissipation from the fields to plasmas.\citep{zeni11c}
Its relevance to the magnetic diffusion process deserves further research. 
In the case of the IDR, it was recently pointed out that
the magnetic diffusion should involve the violation of the ion frozen-in condition and
that the chaotic particle dynamics explains the ion nonidealness
outside the diffusion-like region.\citep{zeni13} 

In this contribution,
we discuss basic properties of the DRs
in line of Ref.~\onlinecite{zeni13} in more detail.
\bluetext{We briefly outline our interpretation of magnetic diffusion,
based on fundamental concepts.\citep{newcomb58,stern66,hornig96,scudder97,priest00}
Particular attention is paid to one case of the violation of the frozen-in condition. } 
Next we test our idea by a Vlasov simulation.
Then, we discuss the relevance between
the DRs and the dissipation region
from the viewpoint of the energy flow,
followed by the summary.

We begin with an arbitrary Ohm's law,
$\vec{E} + \vec{v} \times \vec{B} = \vec{R}$,
where $\vec{v}$ is a smooth velocity field
and $\vec{R}$ is the sum of the nonideal effects.
The induction equation yields
\begin{align}
\label{eq:induction}
{\partial_t} \vec{B}- \rot (\vec{v}\times\vec{B})
+ \rot \vec{R}
= 0.
\end{align}
The last term is introduced by the nonidealness, $\vec{R}\ne 0$.
An Ohm's law $\vec{R}=\eta\vec{j}$ with a simple scalar resistivity yields
\begin{align}
\label{eq:approx}
{\partial_t} \vec{B}- \rot (\vec{v}\times\vec{B})
- \alpha \Delta \vec{B}
= 0,
\end{align}
where $\alpha = \eta/\mu_0$ is the diffusion coefficient. 
This suggests that
the magnetic field relaxes into the $\Delta \vec{B}=0$ state
through magnetic diffusion. 
Although the exact form of $\vec{R}$ is unknown in a kinetic plasma,
we expect that
the $\rot \vec{R}$ term approaches zero
through magnetic diffusion-like processes.

The $\rot \vec{R}=0$ condition is known as
the flux preservation or the flux frozen-in.\citep{newcomb58,stern66} 
When this condition is met,
the total magnetic flux through a surface
which is moving with velocity $\vec{v}$ remains constant,
as illustrated in Fig.~1(a). 
Note that the ideal condition $\vec{R} = 0$ is
sometimes incorrectly referred to as the frozen-in condition. 
The frozen-in condition $\rot \vec{R}=0$
can be violated by either or both of the following ways:
$\vec{b} \cdot (\rot \vec{R}) \ne 0$ and
$\vec{b} \times (\rot \vec{R}) \ne 0$,
where $\vec{b}=\vec{B}/|B|$ is the field-aligned unit vector.
\bluetext{The first case, $\vec{b} \cdot (\rot \vec{R}) \ne 0$,
has largely been ignored in previous literature. } 
It stands for the nonideal magnetic expansion or compression.
Since $\vec{B}$ is divergence-free,
the field lines are transported across the curve that bounds the surface (Fig.~1(b)). 
\bluetext{The second case is related to
the magnetic connectivity of fluid elements,
formally known as the line preservation.\cite{newcomb58,stern66}
We interpret it as the rotation of the field line (Fig.~1(c)). } 
For example, if no magnetic flux initially penetrates a comoving surface
tangent to the field line,
the field line (the dashed line) will penetrate the surface. 
The two fluid elements, initially on the same field line,
will be threaded by different field lines after the rotation. 

We focus on the first case of the violation,
because it better fits our intuition of diffusion processes. 
Let us introduce the loss rate of magnetic flux
due to the nonideal effects,
\begin{align}
\label{eq:loss}
\mathfrak{L}
\equiv
\vec{b} \cdot (\rot \vec{R})
.
\end{align}
\bluetext{This corresponds to a free parameter `$\mu$'
in a topology-conserving flow [$\vec{b} \times (\rot \vec{R}) = 0$] in \citet{hornig96},
i.e., $\mathfrak{L} = \mu |\vec{B}|$. } 
During magnetic diffusion-like processes,
we expect that the magnetic field approaches the $\rot \vec{R}=0$ state
via magnetic loss or gain, $\mathfrak{L} \ne 0$.
The direction of the flux transfer depends on the {context}.
In general it tends to smooth a magnetic profile, as indicated by Eq.~\eqref{eq:approx}. 
To see the context, we compare Eqs.~\eqref{eq:induction} and \eqref{eq:approx}
to estimate an effective diffusivity,
\begin{align}
\label{eq:alpha}
\alpha_{\rm eff}
\equiv -\frac{ \vec{b} \cdot ( \rot \vec{R} )}{\vec{b} \cdot \Delta \vec{B}}
.
\end{align}
We expect $\alpha_{\rm eff}>0$ in a diffusion-like context.
One may further incorporate
the spatial variation of the diffusivity in Eq.~\eqref{eq:approx}.
However,
since we already limit ourselves to the field-aligned component,
and
since the true diffusivity will not be a scalar anyway,\cite{spitzer62,braginskii65}
it is plausible to keep our analysis simple.

Importantly,
the Ohm's law is a function of $\vec{v}$,
that is, $\vec{R}=\vec{R}(\vec{v})$.
Therefore, the idealness, the frozen-in, and magnetic diffusion are
{\itshape relative} concepts that depend on
the reference velocity field $\vec{v}$.
Regarding the velocity field $\vec{v}$,
a magneto-fluid velocity is a natural choice in the MHD,
but we have many choices in a kinetic plasma. 
In the case of the EDR,
it is reasonable to employ the electron velocity $\vec{v}_e$ as a reference.
We expect that
the magnetic field relaxes toward the $\rot\vec{R}(\vec{v}_e)=0$ state
in the EDR. 
We can similarly discuss the IDR by choosing $\vec{v}_i$ as a reference velocity field. 
\begin{figure}[thbp]
\centering
\includegraphics[width={\columnwidth},clip]{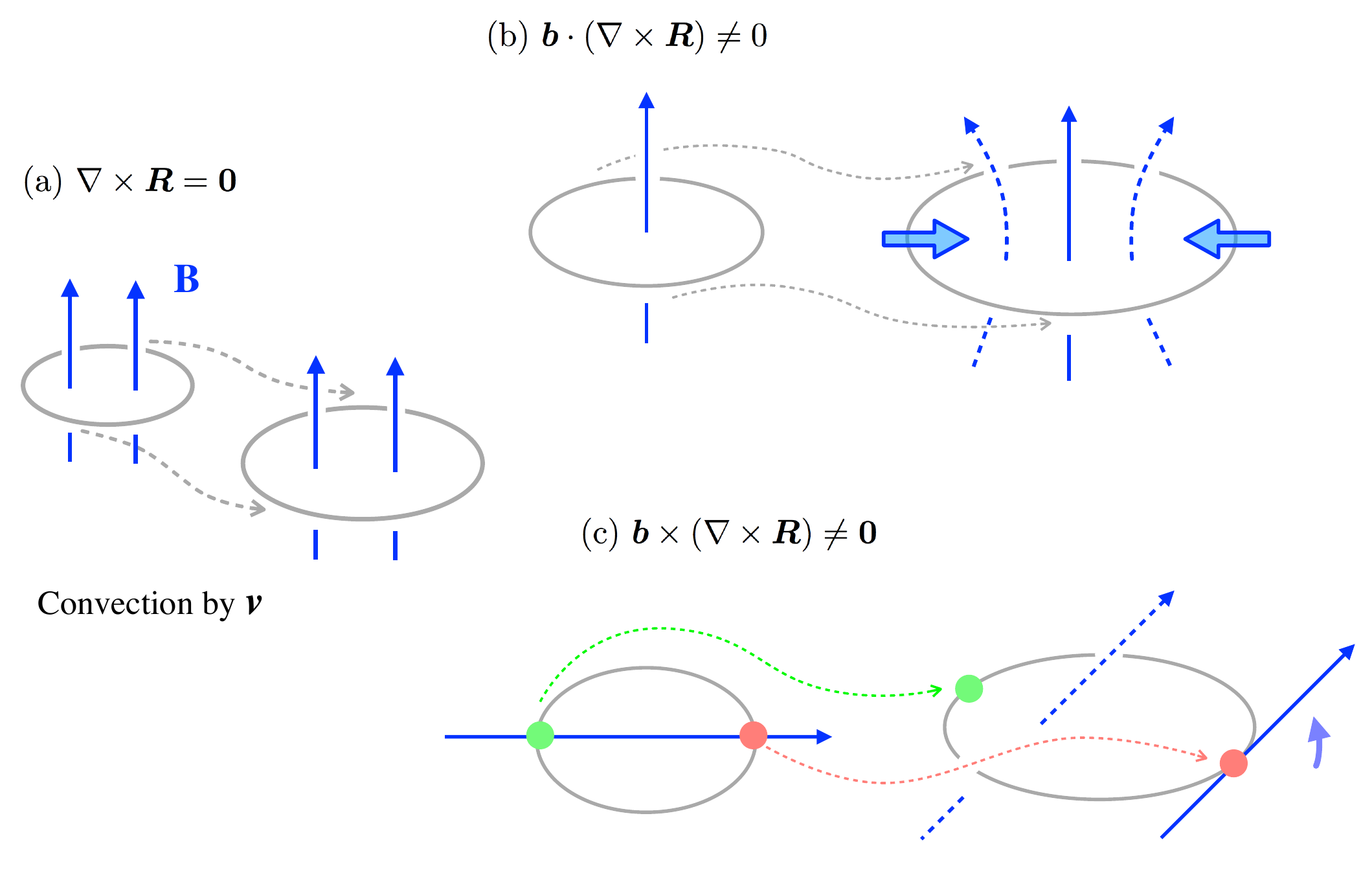}
\caption{
(a) Flux preservation, $\rot \vec{R} = 0$:
the magnetic flux through the curve is conserved.
(b) The violation of the flux preservation in the field-aligned component,
$\vec{b} \cdot (\rot \vec{R}) \ne 0$, and
(c) in the perpendicular components, $\vec{b} \times (\rot \vec{R}) \ne 0$.
}
\label{fig:diagram}
\end{figure}

We carry out an electromagnetic Vlasov simulation of magnetic reconnection
to test the above idea and diffusion measures (Eqs.~\eqref{eq:loss} and \eqref{eq:alpha}).
In contrast to a popular particle-in-cell (PIC) method,
the Vlasov method is free from particle noise and therefore
more suitable to see a basic structure \bluetext{and its spatial gradients. } 
Our code solves two spatial dimensions and three velocity dimensions
($\mathbf{r}\equiv(x,z)$) and ($\mathbf{v}\equiv(v_x,v_y,v_z)$).
It employs a conservative semi-Lagrangian scheme\cite{umeda09}
and includes recent improvements.\cite{umeda12b,umeda12c} 
%
We employ normalized units:
lengths to the ion inertial length $d_i=c/\omega_{pi}$,
times to the inverse ion cyclotron frequency $\Omega_{ci}^{-1}=m_i/(eB_0)$,
velocities to the typical ion Alfv\'{e}n speed $c_{Ai}=B_0/(\mu_0 m_i n_0)^{1/2}$, and
the electric current to $J_0=e n_0 c_{Ai}$.
Here, $\omega_{pi}=(e^2n_0/\varepsilon_0m_i)^{1/2}$ is the ion plasma frequency,
$n_0$ is the reference plasma density, and
$B_0$ is the background magnetic field.
We consider a Harris-type initial model in the $x$-$z$ plane,
$\vec{B}(z)=B_0 \tanh(z/L) \vec{e}_x$ and
$n(z) = n_{0} [0.2 + \cosh^{-2}(z/L)]$,
where $L=0.5 d_i$ is the half thickness of the current sheet.
Ions are assumed to be protons.
The mass ratio is $m_{i}/m_{e}=25$,
the frequency parameter is $\omega_{pe}/\Omega_{ce}=4$, and
the temperature ratio is $T_i/T_e=5$. 
The domain size is $x, z \in[-12.8,12.8]\times[-6.4, 6.4]$.
Considering the symmetry,
we internally solve a quadrant $x, z \in[0,12.8]\times[0, 6.4]$.
The reconnection is triggered at the center $(x,z)=(0,0)$ by a weak perturbation.
Open boundaries are used at the outer boundaries ($x=12.8$ and $z=6.4$).
The quadrant is resolved by $640 \times 320$ grid cells.
The velocity space is resolved by $40^3$ grid cells
with $\Delta v_i=0.29$ and $\Delta v_e=0.72$.

\begin{figure*}[tbp]
\centering
\includegraphics[width={\columnwidth},clip]{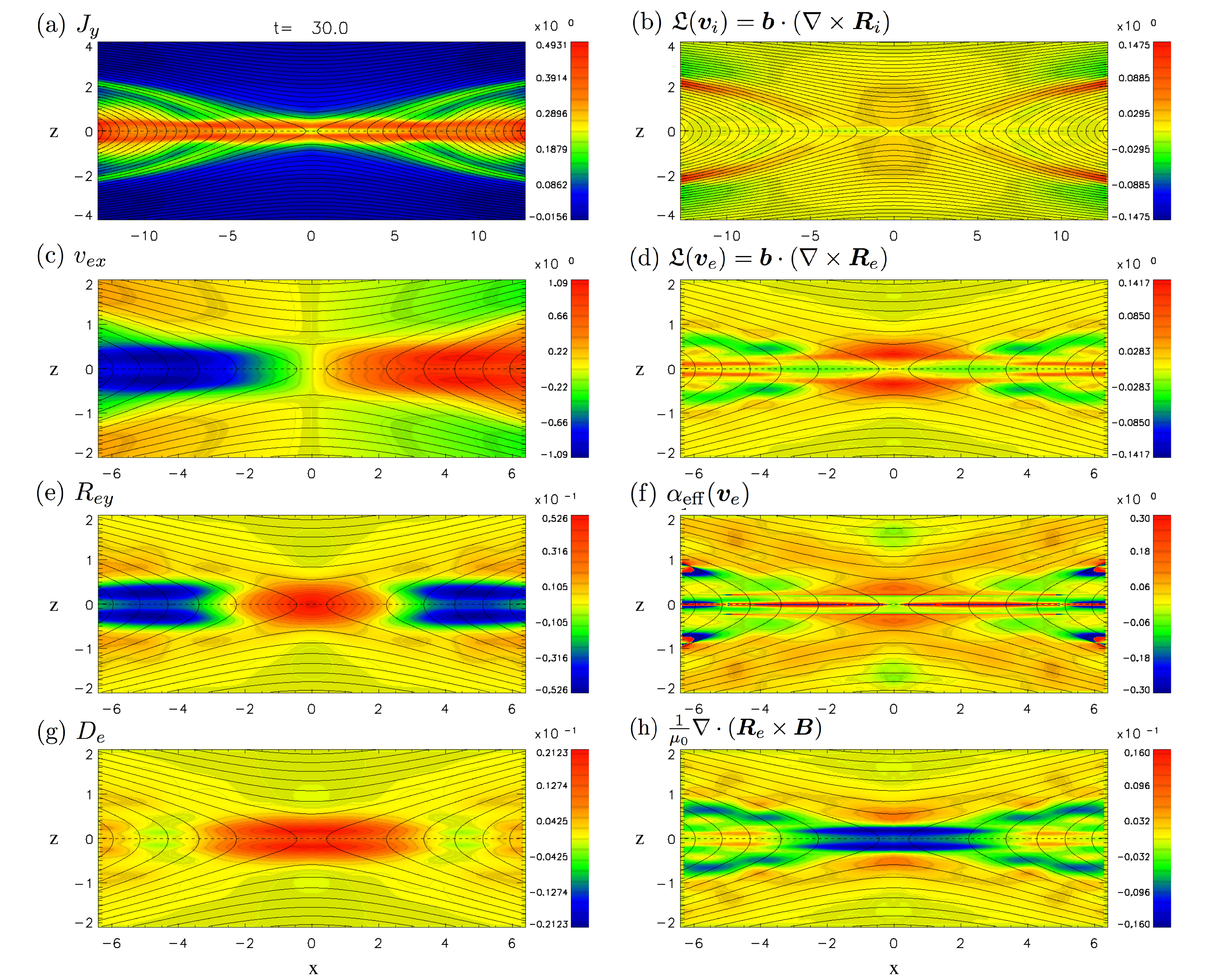}
\caption{
Results of the main run at $t=30$.
The contour lines are in-plane magnetic field lines and
the dashed line indicates the field reversal, $B_x=0$.
(a) The out-of-plane current density $J_y$,
(b) the magnetic loss with respect to the ion flow $\mathfrak{L}(\vec{v}_i)$,
(c) the electron outflow speed $v_{ex}$,
(d) the magnetic loss with respect to the electron flow $\mathfrak{L}(\vec{v}_e)$,
(e) the electron nonideal electric field $R_{ey}$ normalized by $c_{Ai}B_0$,
(f) an effective diffusion coefficient $\alpha_{\rm eff}$ (Eq.~\eqref{eq:alpha}) for $\vec{v}_e$,
(g) the nonideal dissipation measure $D_e$, and
(h) the divergence of the electron nonideal Poynting flux.
}
\label{fig:outflow}
\end{figure*}

Panels in Figure 2 show physical quantities at $t=30$.
Although the system is still evolving,
the electron-scale structure near the X-line is well developed at this time.
The flux transfer rate $E_y/(c_{Ai}B_0)$ at the X-line
(the reconnection rate) reaches a quasi-steady state after $t\gtrsim 20$. 
We confirmed that key features are consistent with
the results in PIC simulations.\cite{shay07,zeni11c} 

Figure 2(a) shows the out-of-plane current density, $J_{y}$. 
It has a two-scale structure\cite{ishizawa04} of
an intense electron current layer near the midplane ($|z| \lesssim 0.5$) and
a broader ion current layer over few $d_i$'s.
Figure 2(b) shows
the magnetic loss with respect to the ion flow,
$\mathfrak{L}(\vec{v}_i)$.
The positive region ranges widely over $|x| \lesssim 7$.
Since the magnetic field decreases from the upstream regions to the midplane,
and since the magnetic gradient (the electric current) becomes steeper
in the closer vicinity of the midplane ($z=0$),
such magnetic loss is consistent with magnetic diffusion.
It is also evident along the separatrices.
This is probably due to the polarization electric field $E_z$, but
we focus on the central DRs in this paper. 

Figure 2(c) focuses on electron outflow $v_{ex}$
in the center of the simulation domain. 
Narrow bidirectional jets from the X-line are
well-known signatures of kinetic reconnection.\cite{shay07}
They also correspond to the electron current layer in $|z| \lesssim 0.5$ (Fig.~2(a)). 
Electrons undergo Speiser-type bounce motion inside this layer.\citep{speiser65} 
Figure 2(d) shows the magnetic loss with respect to the electron flow,
$\mathfrak{L}(\vec{v}_e)$.
In the inflow direction,
the magnetic field nonideally decreases
around the compact (red) regions, $|x| \lesssim 2.5$.
The magnetic field is nonideally compressed
in the outflow direction along $z=0$, as indicated in green.
These are expected signatures of the DR.
In addition, there are narrow compression regions
on the boundaries of the electron jets around $|x| > 3, |z| \approx 0.5$.
This anti-diffusion signature is less apparent in PIC simulations. 
In Fig.~2(e), we show the out-of-plane component of the nonideal electric field,
$R_{ey} = (\vec{E}+\vec{v}_e\times\vec{B})_y$.
It consists of the central $R_{ey}>0$ region and
the outer electron jets with $R_{ey}<0$,
as previously reported.\cite{shay07}
Since $\mathfrak{L}(\vec{v}_e) \approx \pm\partial_z R_{ey}$ around the center,
the length of the loss regions (Fig.~2(d)) are
comparable with the length of the $R_{ey}>0$ region.
Meanwhile,
the compressional (green) region enters the $R_{ey}<0$ region at $z=0$,
as long as $\mathfrak{L}(\vec{v}_e) \approx \pm \partial_x R_{ey}$ is nonzero. 
Figure 2(f) shows an effective diffusivity
with respect to the electron velocity, $\alpha_{\rm eff}(\vec{v}_e)$
(Eq.~\eqref{eq:alpha}).
We recognize the $\alpha_{\rm eff} > 0$ region near the center.
This indicates that
the magnetic flux relaxes to the electron frozen-in state
in a diffusion-like manner. 
It looks puzzling near the midplane $|z|<0.25$,
because the electric current density has two peaks at $z \approx \pm 0.25$ (Fig.~2(a)).
Here electrons are bouncing in $z$ inside the electron current layer.
It is possible that our simple analysis is no longer useful
in such an extreme kinetic limit, and so
we may better to rule out this narrow layer. 
Summarizing these results, 
although there is some ambiguity,
we recognize a compact region
around $|x| \lesssim 3, |z| \lesssim 0.7$ (Figs.~2(d) and 2(f)).
Since this is identified by a diffusion-like behavior of
the magnetic field with respect to the electron flow,
it is reasonable to call it the EDR.
As expected, the EDR is an electron-scale localized region near the X-line.


Figure 2(g) shows a measure of the nonideal energy transfer,\cite{zeni11c}
$D_e \equiv \vec{j}\cdot(\vec{E}+\vec{v}_e\times\vec{B})-\rho_c\vec{v}_e\cdot\vec{E}$,
where $\rho_c$ is the charge density.
Previous research demonstrated that
$D_e$ marks a compact site of energy dissipation near the X-line,
i.e., the dissipation region.\citep{zeni11c}
The EDR looks similar to, but
slightly wider in $z$ and shorter in $x$
than the dissipation region. 
We attribute this minor difference to the energy flow. 
Applying $\vec{B}\cdot$ to Eq.~\eqref{eq:induction},
we obtain the nonideal part of the magnetic energy equation,
\begin{align}
\label{eq:energy}
\frac{1}{\mu_0} \vec{B} \cdot (\rot \vec{R})
=
\frac{1}{\mu_0}\div (\vec{R} \times \vec{B} )
+
\vec{j} \cdot \vec{R}
.
\end{align}
For $\vec{v}=\vec{v}_e$,
the left hand side gives a similar picture as Fig.~2(d).
\bluetext{The first term on the right is presented in Fig.~2(h)
with respect to $\vec{v}_e$.
We interpret $\frac{1}{\mu_0}\vec{R}\times\vec{B}$
as the nonideal part of Poynting flux,
because we can split Poynting flux into
$\frac{1}{\mu_0}\vec{E}\times\vec{B}=\frac{1}{\mu_0}(-\vec{v}\times\vec{B})\times\vec{B}+\frac{1}{\mu_0}\vec{R}\times\vec{B}$.
This nonideal Poynting flux is equivalent to Poynting flux observed
in the moving frame with $\vec{v}$. 
The divergence term in Eq.~\eqref{eq:energy} stands for
the energy transfer by the nonideal Poynting flux. 
} 
The last term corresponds to the nonideal energy transfer
between the field and plasmas (Fig.~2(g)),
as $D_e = \vec{j} \cdot \vec{R}(\vec{v}_e) = \vec{j} \cdot \vec{R}(\vec{v}_i)$
in a neutral ion-electron plasma. 

On the inflow side of the EDR,
the magnetic field departs from the electron flow (Fig.~2(d)).
This is a magnetic diffusion-like behavior. 
\bluetext{As the electric field becomes nonideal,
Poynting flux is carried by its nonideal part (${\approx}-\frac{1}{\mu_0}R_{ey}B_x$). }
Thus the nonideal Poynting flux emanates here (Fig.~2(h)). 
It carries energy in $z$ and then
disappears near the midplane (the blue sinks in Fig.~2(h)),
where a plasma gains energy instead, as marked by $D_e$ (Fig.~2(g)). 
As the electrons diverge in $x$ through the Speiser-type motion,
the sink region ranges longer in $x$ than the source regions of
the electron nonideal Poyting flux. 
In short, the EDR and the dissipation region can be regarded as
the upstream source or the downstream sink of the nonideal Poynting flux. 

Diffusion signatures are also found
with respect to the ion flow (Fig.~2(b)).
This region should include the IDR. 
At this point, the size of the IDR remains unclear,
because our simulation box is too small. 
From the energy viewpoint,
the difference between the IDR and EDR arises from
the Poynting flux by the Hall electric field in the generalized Ohm's law, 
\begin{align}
\frac{B}{\mu_0}
\Big(
\mathfrak{L}(\vec{v}_i)
-
\mathfrak{L}(\vec{v}_e)
\Big)
=
\div \Big(\frac{\vec{j}\times\vec{B}}{en} \times \vec{B} \Big)
.
\end{align}
This makes sense, because the Hall physics should play a role in the IDR. 
As well known, the Hall electric field does not involve
direct energy conversion to plasmas. 
This is complement to the fact
that the IDR does not involve significant energy dissipation $D_e$. 
Practically, it is difficult to evaluate
the diffusivity (Eq.~\eqref{eq:alpha}) for the ion flow,
because $\Delta \vec{B}$ already contains electron-scale structures.
For better discussion,
it is necessary to evaluate magnetic diffusion in a coarse manner,
by averaging physical quantities over an ion-physics scale. 

\bluetext{
We have examined magnetic diffusion properties
in one configuration: symmetric anti-parallel reconnection. 
Results can be different in guide-field configuration,
in which a non-reconnecting magnetic field dominates around the reconnection point. 
Moreover, our diffusion properties differ from the magnetic tilt, or
the violation of the line preservation [$\vec{b}\times (\rot\vec{R})\ne 0$],
which indicates ``reconnection'' of the field lines. 
In order to see whether magnetic diffusion is essential,
the line preservation as well as the diffusion properties
need to be investigated in various types of magnetic reconnection. 
} 



In summary, we have studied the diffusion regions
surrounding the X-line in magnetic reconnection.
\bluetext{Recognizing the role of the frozen-in condition in magnetic diffusion,
we have proposed the diffusion properties and 
have applied them to anti-parallel magnetic reconnection.
Taking advantage of the low-noise Vlasov simulation, 
we have recognized the electron-scale and ion-scale DRs. } 
The EDR is almost similar to the dissipation region.
The minor difference is attributed to the nonideal part of Poynting flux near the X-line.
The difference to the ion-scale region is attributed to Hall physics. 
These results are consistent with a previous expectation\cite{hesse01}
that the electron-scale region of the energy dissipation is surrounded
by the larger Hall-physics region. 



\begin{acknowledgments}
This research was supported by Grant-in-Aid (KAKENHI)
No.25871054 (SZ) and No.25610144 (TU).
Computation was performed on CX400 at Kyushu Univ.
as HPCI/JHPCN programs (hp120165 and jh130005-NA03).
\end{acknowledgments}

%


\begin{thebibliography}{}

\bibitem[Burch \& Drake(2009)]{burch09}
J. L. Burch and J. F. Drake, American Scientist {\bf 97}, 392 (2009).
\bibitem[Wilmot-Smith et al.(2005)]{wilmotsmith05}
A. L. Wilmot-Smith, E. R. Priest, \& G. Hornig,
Geophysical \& Astrophysical Fluid Dynamics {\bf 99}, 177 (2005).

\bibitem[Biskamp et al.(1997)]{biskamp97}
D. Biskamp, E. Schwarz, and J. F. Drake, \pop {\bf 4}, 1002 (1997).
\bibitem[Hesse et al.(2001)]{hesse01}
M. Hesse, J. Birn, and M. Kuznetsova, \jgr {\bf 106}, 3721 (2001).
\bibitem[Drake \& Shay(2007)]{drake07}
J. F. Drake and M. A. Shay, in
``Reconnection of Magnetic Fields: Magnetohydrodynamics and Collisionless Theory and Observations,''
edited by J. Birn and E. R. Priest
(Cambridge University Press, Cambridge, 2007), Sec. 3.1.

\bibitem[Mozer(2005)]{mozer05}
F. S. Mozer, \jgr {\bf 110}, A12222, doi:10.1029/2005JA011258 (2005).
\bibitem[Shay et al.(2007)]{shay07}
M. A. Shay, J. F. Drake, and M. Swisdak, \prl {\bf 99}, 155002 (2007).
\bibitem[Scudder \& Daughton(2008)]{scudder08a}
J. Scudder and W. Daughton, \jgr {\bf 113}, A06222, doi:10.1029/2008JA013035 (2008).
\bibitem[Pritchett \& Mozer(2009)]{prit09a}
P. L. Pritchett and F. S. Mozer, \pop {\bf 16}, 080702 (2009).
\bibitem[Zenitani et al.(2011a)]{zeni11c}
S. Zenitani, M. Hesse, A. Klimas, and M. Kuznetsova, \prl {\bf 106}, 195003 (2011).
\bibitem[Zenitani et al.(2013)]{zeni13}
S. Zenitani, I. Shinohara, T. Nagai, and T. Wada, \pop {\bf 20}, 092120 (2013).

\bibitem[Newcomb(1958)]{newcomb58}
W. A. Newcomb, Ann. Phys. {\bf 3}, 347 (1958).
\bluetext{
\bibitem[Stern(1966)]{stern66}
D. P. Stern, \ssr {\bf 6}, 147 (1966).
\bibitem[Hornig \& Schindler(1996)]{hornig96}
G. Hornig and K. Schindler, \pop {\bf 3}, 781 (1996).
\bibitem[Scudder(1997)]{scudder97}
J. D. Scudder, \ssr {\bf 80}, 235 (1997).}
\bibitem[Priest and Forbes(2000)]{priest00}
E. Priest and T. Forbes,
``Magnetic reconnection : MHD theory and applications'', New York : Cambridge University Press (2000), Chaps. 1 and 8.
\bibitem[Spitzer(1962)]{spitzer62}
L. Spitzer, Jr., {\it Physics of Fully Ionized Gases}, Interscience, New York, (1962).
\bibitem[Braginskii(1965)]{braginskii65}
S. I. Braginskii, Rev. Plasma Phys. {\bf 1}, 205 (1965).

\bibitem[Umeda et al.(2009)]{umeda09}
T. Umeda, K. Togano, and T. Ogino,
Comput. Phys. Commun. \textbf{180}, 365 (2009).
\bibitem[Umeda et al.(2012)]{umeda12b}
T. Umeda, Y. Nariyuki, and D. Kariya,
Comput. Phys. Commun. \textbf{183}, 1094 (2012).
\bibitem[Umeda et al.(2012)]{umeda12c}
T. Umeda, K, Fukazawa, Y. Nariyuki, and T. Ogino,
IEEE Trans. Plasma Sci. \textbf{40}, 1421 (2012).

\bibitem[Ishizawa et al.(2004)]{ishizawa04}
A. Ishizawa, R. Horiuchi, and H. Ohtani, \pop {\bf 11}, 3579 (2004).
\bibitem[Speiser et al.(1965)]{speiser65}
T. W. Speiser, \jgr {\bf 70}, 4219, doi:10.1029/JZ070i017p04219 (1965).

\end{thebibliography}
\end{document}